\documentclass[12pt]{article}
\usepackage{graphicx}
\begin{document}

\centerline{\bf Inbreeding and outbreeding depressions  in the}

\centerline{\bf Penna model as a result of crossover frequency}
\bigskip

\centerline{K. Bo\'nkowska, M. Kula, S. Cebrat and D. Stauffer$^*$}

\bigskip
\noindent
Department of Genomics, 
Wroc{\l}aw University, ul. Przybyszewskiego 63/77, 51-148 Wroc{\l}aw, Poland

\medskip
\noindent
$^*$ Visiting from Institute for Theoretical Physics, Cologne University, 
D-50923 K\"oln, Euroland

\bigskip
Abstract: The population in the sexual Penna ageing model is first separated into 
several reproductively isolated groups. Then, after equilibration, sexual mixing
between the groups is allowed. We study the changes in the population size due 
to this mixing and interpret them through a counterplay of purifying selection 
and of haplotype complementarity.
\bigskip

\section{Introduction}
In sexual reproduction, as opposed to asexual reproduction, the genomes of the 
two parents are mixed, and within the diploid genome of each parent happens
crossover. This way of reproduction has advantages as well as disadvantages 
compared with asexual cloning of haploid genomes. An advantage is that bad 
recessive mutations do not affect the health if they are present in only one of 
the two haplotypes (= sets of genetic information). A disadvantage
is the reduced number of births if only the females produce offspring while the 
males consume as much food and space as the females. Moreover, crossover of 
two different genomes may produce a mixture which is fitter than each the two 
parents but also one which is less fit, as seen these 
days in the DaimlerChrysler car company (outbreeding depression). For small 
populations, the probability is higher that the two parents have the same bad 
recessive mutation which therefore diminishes the health of the individual 
(inbreeding depression). 

\begin{figure}[hbt]
\begin{center}
\includegraphics[angle=-90,scale=0.5]{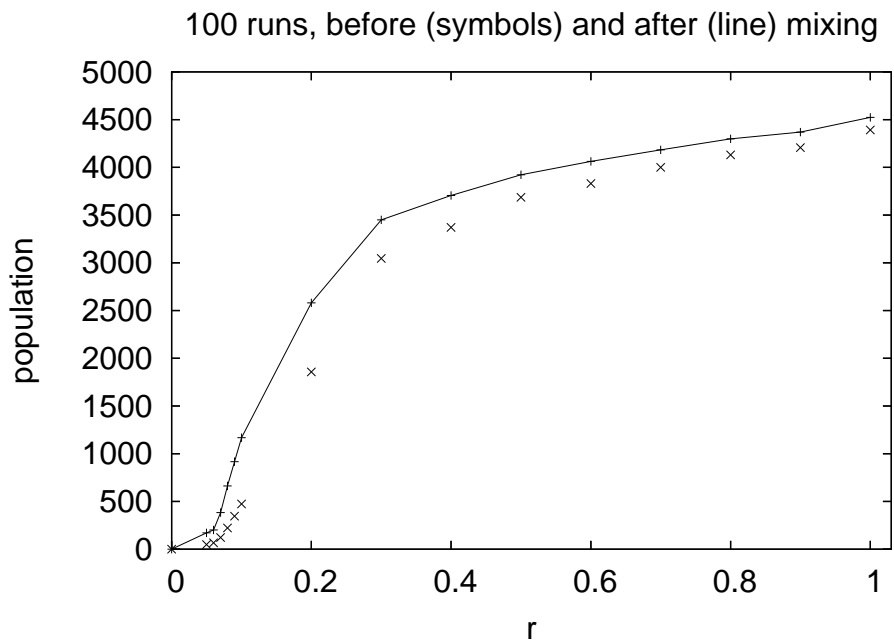}
\end{center}
\caption{Average over 100 simulations with $G=10$ groups each, $\Delta = 100$.
}
\end{figure}

\begin{figure}[hbt]
\begin{center}
\includegraphics[angle=-90,scale=0.5]{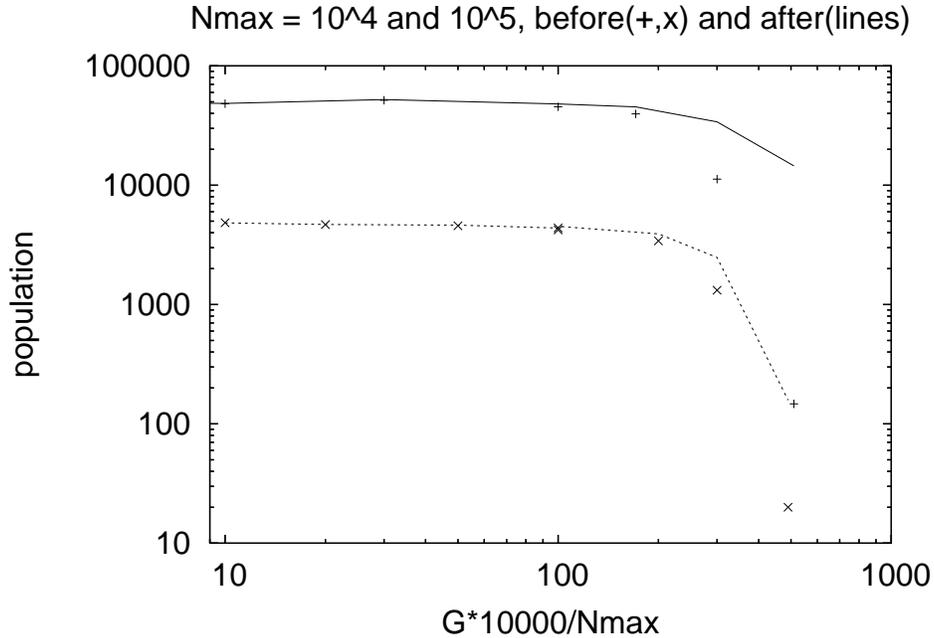}
\end{center}
\caption{Average over 10 simulations with a large (top) and 100 with a small
(bottom) population, versus number $G$ of groups; $r=1, \; \Delta = 1000$ and 
100, respectively. For the larger population $G$ is divided by 10 so that 
data for the same number of individuals per group have the same horizontal
coordinate. We see nice scaling.
}
\end{figure}

\begin{figure}[hbt]
\begin{center}
\includegraphics[angle=-90,scale=0.30]{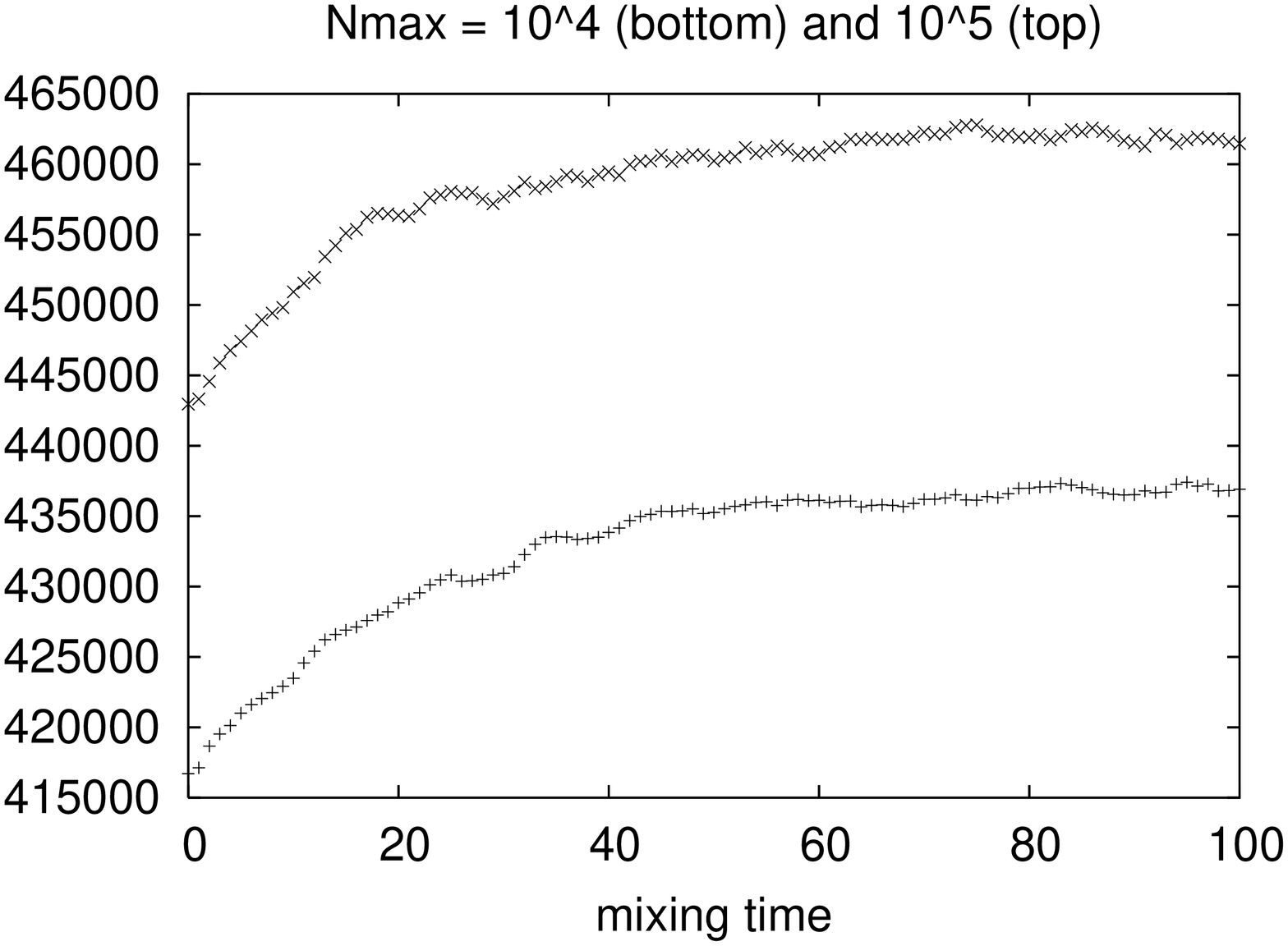}
\includegraphics[angle=-90,scale=0.30]{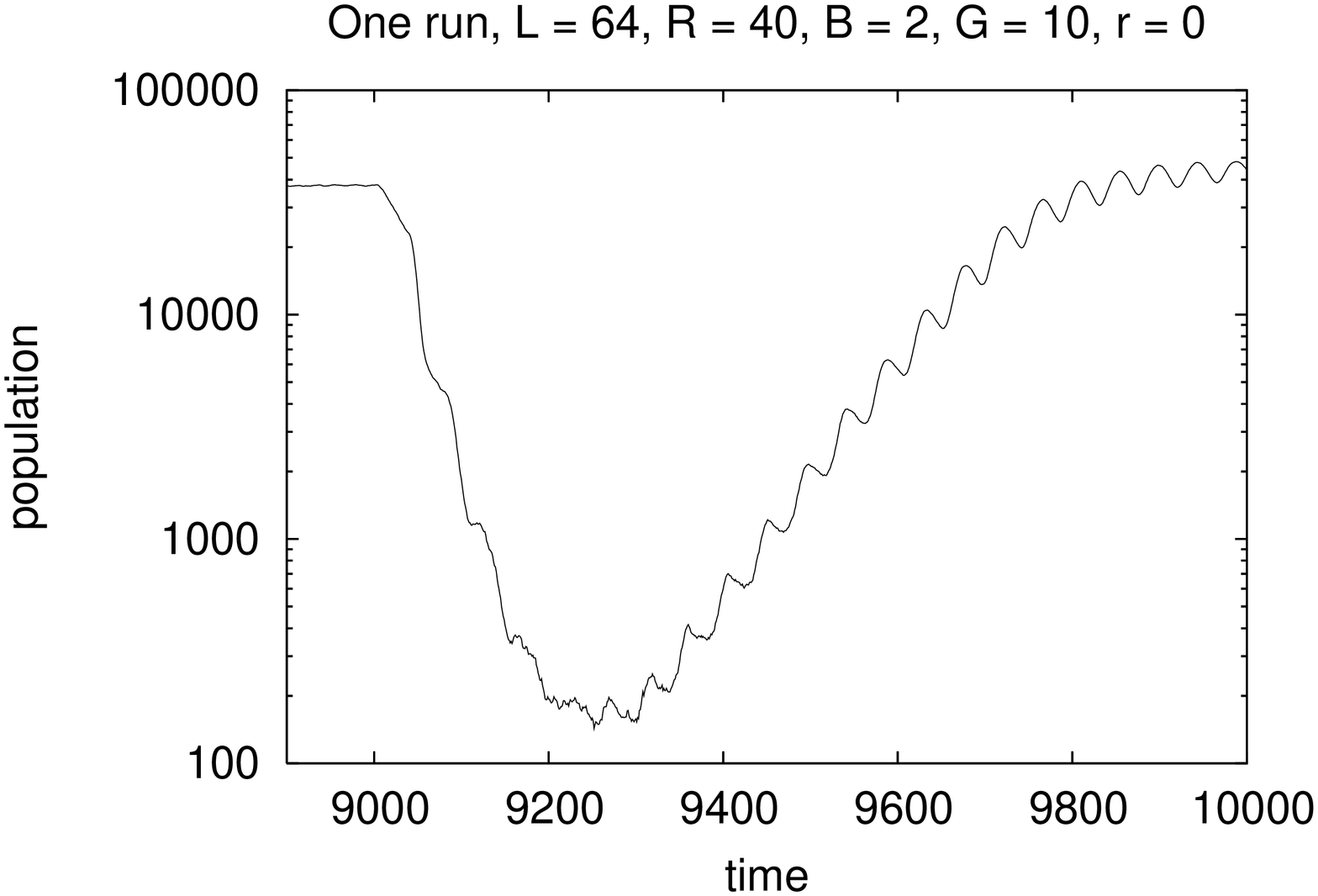}
\end{center}
\caption{Time dependence of outbreeding advantage (part a) and outbreeding
depression (part b).
}
\end{figure}

\begin{figure}[hbt]
\begin{center}
\includegraphics[angle=-90,scale=0.5]{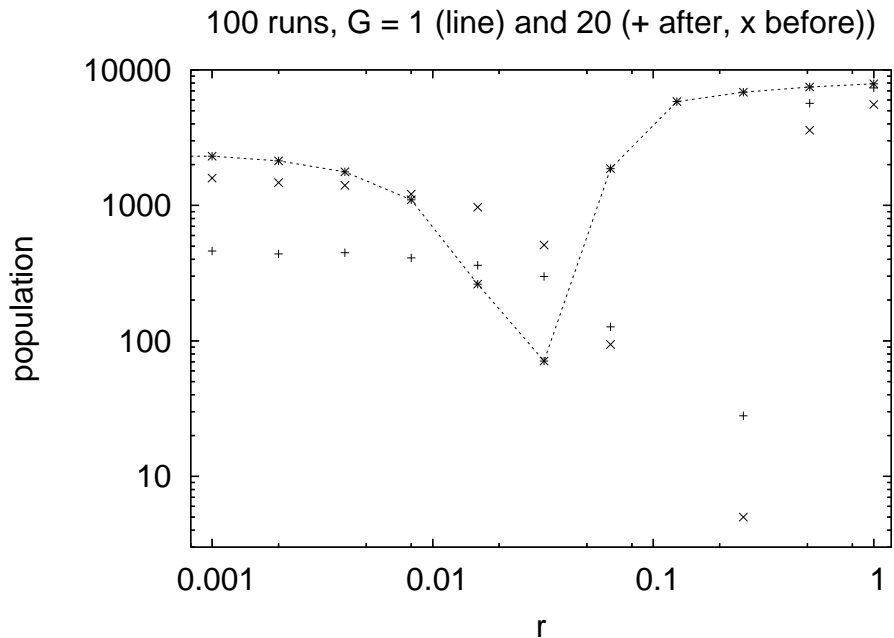}
\end{center}
\caption{Average over 100 simulations with a small population, high minimum
age of reproduction, and $G=1$ and 50. For $G=1$ there is always complete 
mixing. Note the double-logarithmic scales, also in Fig.5 and 6.
}
\end{figure}

\begin{figure}[hbt]
\begin{center}
\includegraphics[angle=-90,scale=0.5]{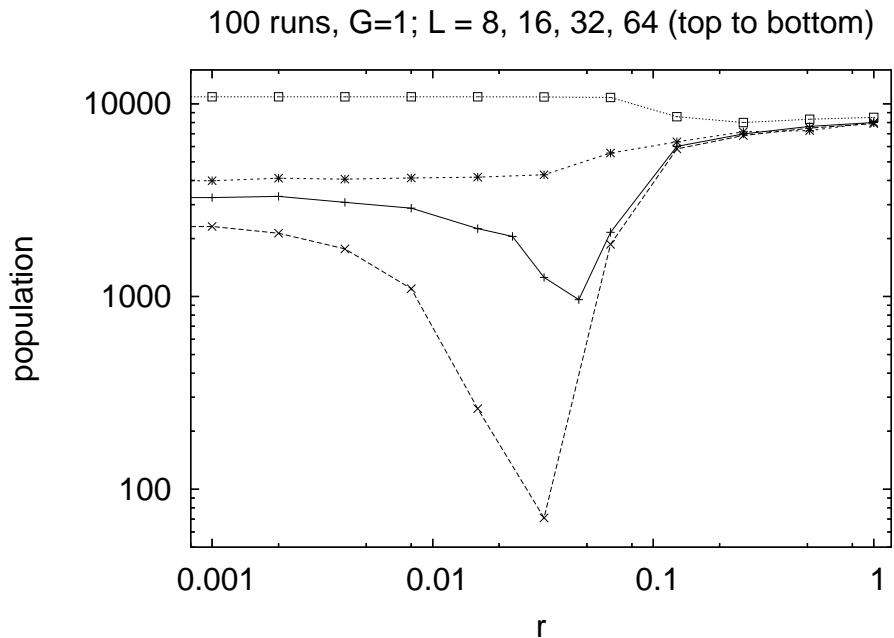}
\end{center}
\caption{Average over 100 simulations with a small population, high minimum
age of reproduction and various
lengths $L$ of the bit-strings, using a birth rate $B = 128/L$. 
}
\end{figure}

\begin{figure}[hbt]
\begin{center}
\includegraphics[angle=-90,scale=0.5]{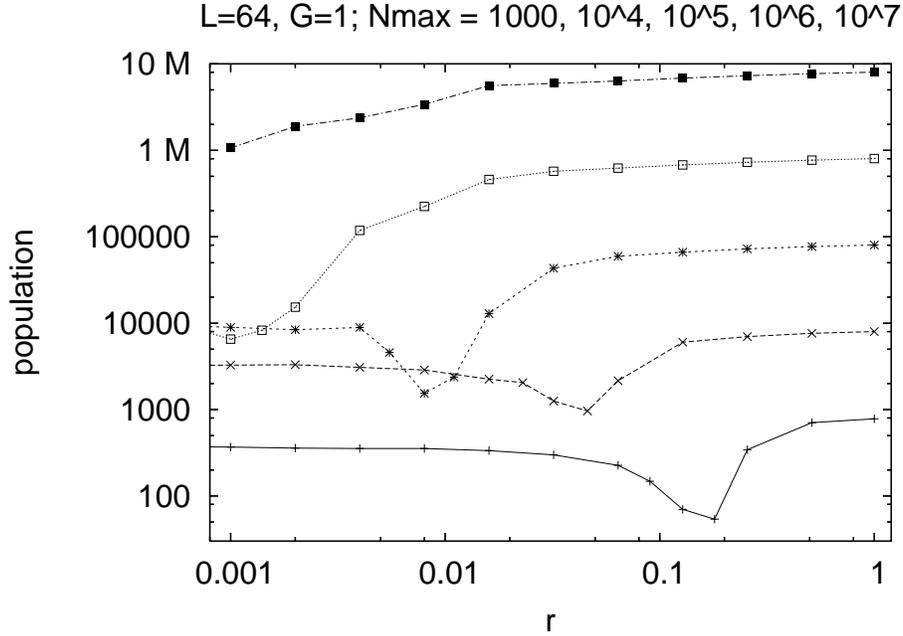}
\end{center}
\caption{Dependence on population size for $N_{\max} = 10^3 \dots 10^7$, 
averaged over 1000 to one sample. $L=64, \; B = 2$.
}
\end{figure}

\begin{figure}[hbt]
\begin{center}
\includegraphics[angle=-90,scale=0.5]{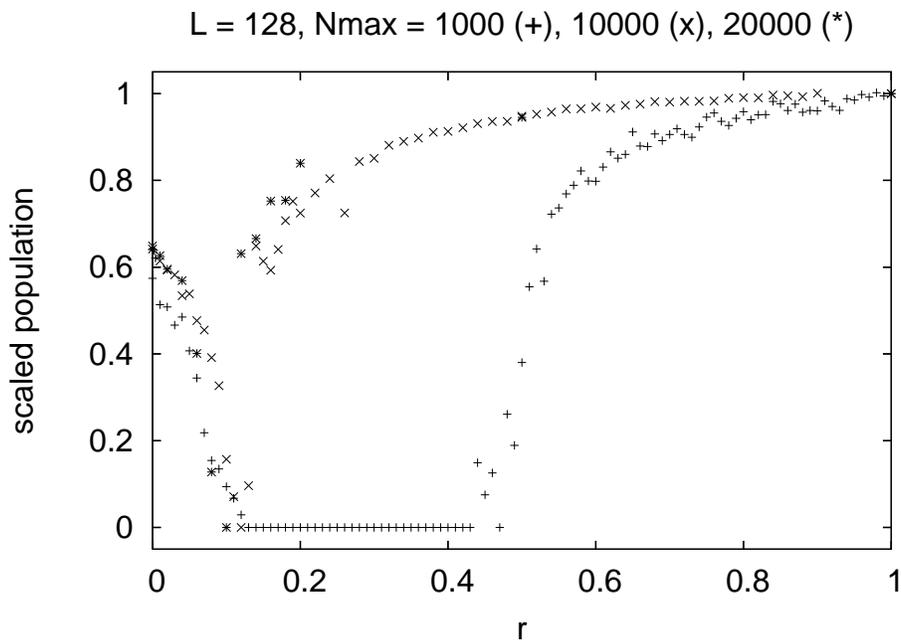}
\end{center}
\caption{
Relation between normalized population size and the crossover
rate. The population size was divided by the population evolving with crossover
rate $r = 1$.
}
\end{figure}

\begin{figure}[hbt]
\begin{center}
\includegraphics[angle=-90,scale=0.5]{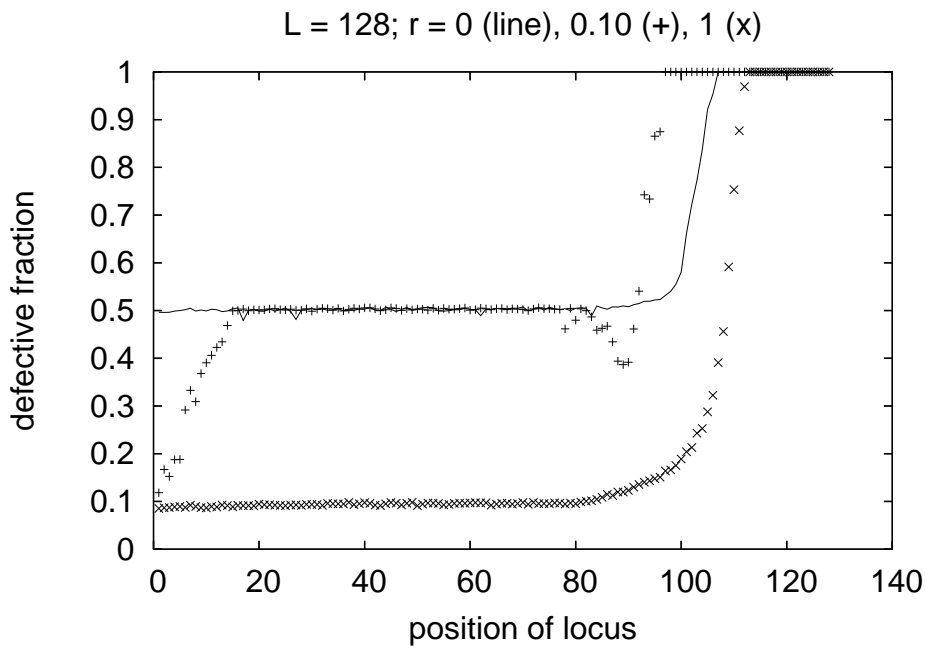}
\end{center}
\caption{
Distribution of defective genes in the genomes of populations
evolving under different crossover frequencies.
}
\end{figure}

\begin{figure}[hbt]
\begin{center}
\includegraphics[angle=-90,scale=0.5]{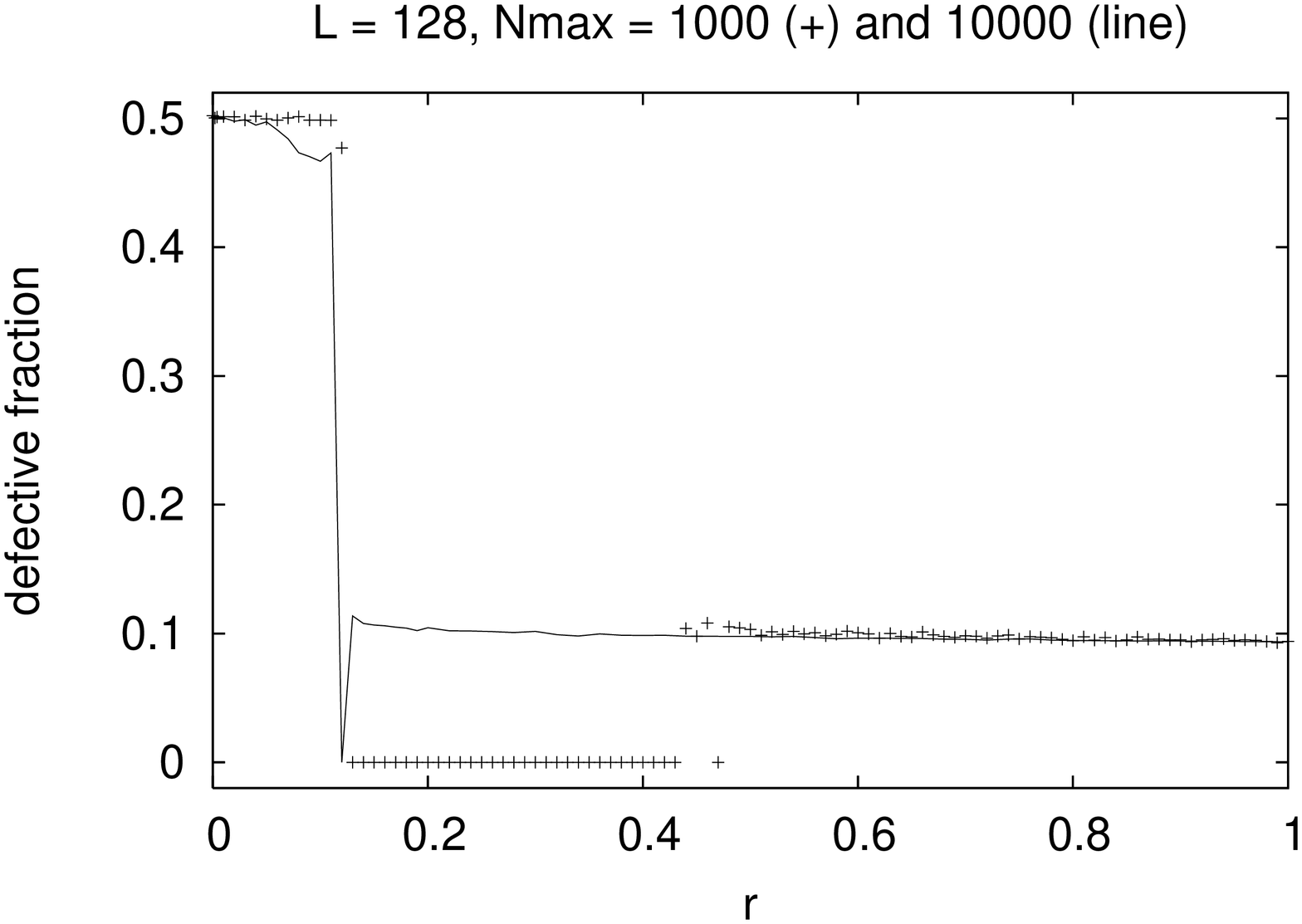}
\end{center}
\caption{
Relation between crossover frequency and average frequency of
defective genes in the sections of genomes expressed before the
reproduction age $R$. Note, fractions equal 0 mean that the populations with 
the given crossover frequency died out.
}
\end{figure}

\section{Standard Model}

We try to simulate these effects in the standard sexual Penna ageing model,
deviating from published programs \cite{books} as follows: The Verhulst factor, 
a death probability $N/N_{\max}$ at population $N$ with carrying capacity
$N_{\max} = 10^3, \ 10^4, \,10^5$,
due to limited food and space, was applied to the births only and not to adults;
the initial age distribution was taken as random when needed;
the birth rate was reduced from 4 to 1, the lethal threshold of active 
mutations from 3 to 1 (that means a single active mutation kills), and mostly 
only $10^4$ instead of $2 \times 10^4$ time steps were made. (One time step or
iteration is one Monte Carlo step for each individual). Furthermore
the whole population was for most of the simulated time separated into $G$ 
different groups such that females look for male partners only within their
own group, with a separate Verhulst factor applying to each group. For the
last $\Delta \ll 10^4$ time steps this separation into groups was dissolved: 
Then females could select any male, and only one overall Verhulst factor 
applied to the whole population. Finally, the crossover
process within each parent before each birth was not made always but only
with a crossover probability $r$. 

If there would be no inbreeding depression then during the first longer part
of the simulation the total number $N_1$ of individuals would be independent of 
the number $G$ of groups into which it is divided. And if then there are no 
advantages or disadvantages of outbreeding, the population $N_2$ during the 
second shorter part, $10^4 - \Delta < t < 10^4$, would be the same as the
preceding population $N_1$ during the last section, $10^4 - 2 \Delta < t < 
10^4 - \Delta$, of the longer first part. We will present data showing that 
this is not the case. Similar simulations for $G=2$ groups was published
long ago \cite{LSC}. 
 
A difficulty in such simulations is the Eve effect: After a time 
proportional to the population size, everybody has the same female (Eve) and 
the same male (Adam) as ancestor, with all other offspring having gotten less 
fit genomes due to random mutations and thus having died out. If we would
divide the whole population into many groups without further changes, the Eve 
effect would let all groups but one die out and thus destroy the separation.
Therefore for the first long period of separation we used separate Verhulst 
factors for each group, stabilizing its population, while for the second shorter
part of mixing we used mostly $\Delta = N_{\max}/100$. 

Figure 1 shows the dependence on the crossover probability for the 
populations $N_1$ before and $N_2$ after mixing. We see that the mixing always
increases the population, that means one has no outbreeding depression but
an outbreeding advantage. Figure 2 confirms this advantage but also 
shows the inbreeding depression: The larger the number $G$ of groups (and 
thus the smaller the group size) is, the smaller are the two populations
$N_1$ and $N_2$. (The difference between $N_1$ and $N_2$ fluctuates less than
these numbers themselves since $N_2$ is strongly correlated with $N_1$.)
Also, for the larger population in Fig.2, the number of groups can be larger
before the population becomes extinct. Figure 3 shows the time dependence 
of the outbreeding effect with mixing between groups allowed after 9900 (part a)
and 9000 (part b) time steps. Figure 3a shows summed populations from 
100 simulations with a small population ($G = 10$) and 10 simulations of a large
population ($G = 100$), versus time after mixing started; $r = 1$ in both cases.
For much larger populations of 5 million and still $G = 10$, no such effect of 
mixing is seen. Part b shows for the high
reproduction age $R$ of the following figures one example of the outbreeding
depression (bottleneck \cite{malarz}) followed by a recovery with oscillations 
of period $R$ after mixing was allowed from time 9001 on; $N_{\max}=10^5$..

We also checked for the influence of $r$ in the case when the minimum age of 
reproduction $R$ is 5/8 of the length $L$ of the bit-strings, i.e. larger than 
the value of 8 used before, and when $L$ is different from the 32 used in Figs.
1 to 3. In these simulations we also assumed all mutations to be
recessive, in contrast to the 6 out of 32 dominant bit positions for Figs.
1 to 3. Figure 4 shows for $L = 32$ and a birth rate $B = 4$ 
a minimum of the population at intermediate $r$ for one group, and for 50 
groups a monotonic behaviour but with outbreeding depression at small $r$ and
outbreeding advantage at big $r$.  This population minimum is seen for
$L = 64$ and 32 but not for 16, Fig.5. Figure 6 shows the dependence on 
population size. (Our data before and after mixing are average over $\Delta = 
100$ or 1000 iterations. When outbreeding depression occurs it may happen that 
later the population recovers: Fig.3b.)

\section{Interpretation}

To study the inbreeding and outbreeding depressions in detail we have
analyzed the results of simulations of single populations of different size
under different regime of intragenomic recombinations (crossover rate $r$).
Parameters for these simulations have been slightly changed to get clearer 
results: $L= 128,\; R = 80,\; N_{\max} = 1000$ to 20 000, crossover probability 
$r =  0$ to 1, $B = 1$, time of simulations = $5 \times 10^5$ iterations.
In Fig. 7 the relation between the size of population and the crossover
probability for three different environment capacities are shown.

Populations in the smallest environment ($N_{\max}=1000$) survive with $r = 0$ 
but their sizes decrease with increasing $r$ and are extinct for $r$ set between
0.12 and 0.4. Under larger crossover rates populations survive and their
sizes are larger than those obtained for $r = 0$ (see plots in Fig. 7 where
sizes of populations were normalized by the size of population under 
$r=1$). Larger populations ($N_{\max}= 10 000$) are extinct in a very
narrow range of crossover rates close to 0.12, and populations with $N_{\max}= 
20000$ become extinct at slightly lower crossover rates. Nevertheless, all
populations have larger sizes when the crossover rate is of the order of 1
per gamete production (the highest tested).

This nonlinear relation between size of population and crossover rate could
be explained on the basis of the genetic structure of individual genomes in
the simulated populations. In Fig. 8 we have shown the frequency of
defective genes in the genetic pool of populations for $N_{\max} = 10000$ under
crossover rates 0, 0.1 and 1. The frequency of defective genes expressed
before minimum reproduction age ($R = 80$) in populations without crossover is
0.5. Since $T = 1$, if the distribution of defects would be random the
probability of any individual to survive until the reproduction age $R$ would be
$0.75^R$ (negligibly small for large $R > 30$). Thus, to survive, individuals 
have to complete their
genomes of two complementing bit-strings (haplotypes). For more efficient
reproduction the number of different haplotypes should be restricted and in
fact there are only two different complementing haplotypes in the whole
population as it was shown in \cite{waga}. In such populations, the
probability of forming the offspring surviving until the reproduction age is
0.5. Note, that recombination at any point inside the part of the genome
expressed before reproduction age $R$ produces a gamete which is not
complementary to any other gamete produced without recombination or with
recombination in an other point. Thus, crossovers in such populations are
deleterious for the offspring.
On the other extreme, with crossover probability = 1, populations are under
purifying selection. The fraction of defective genes in the population is
kept low (about 0.1, compared with 0.5 without recombination), to enable
the surviving of the offspring until their reproduction period. The critical
crossover frequency close to 0.12 is connected with a sharp transition from
these two strategies of genomic evolution: complementarity and purifying
selection. In Fig. 9 the frequency of defective genes expressed before the
reproduction age is plotted. For lower crossover rates the fractions of
defective genes are kept at the level 0.5, for higher crossover rates they
are close to 0.1. Close to the critical frequency of crossover, defective
genes located at both ends of the region of genomes expressed before the
reproduction age are forced to obey the purifying selection which eliminates
some defects (Fig. 8).

In the case of small populations, the probability of meeting two closely related
partners (high inbreeding coefficient) is high and as a consequence, there
is higher probability of meeting two defective alleles in the same locus in
zygote which determines phenotypic defect and eliminates the offspring from
the population. In such condition the strategy of completing the genome of
two complementing haplotypes is more effective. Nevertheless, this strategy
is not the best if effective populations are very large, with low inbreeding
coefficient, when the probability of meeting two identical haplotypes is
negligible. Thus, comparing very large populations with very small ones we
can observe the inbreeding depression. On the other hand, this strategy in
small populations leads to the emerging of very limited number of different
haplotypes in the populations (in extreme only two). These haplotypes are
characterized by a specific sequence of defective alleles. Independent
simulations generate haplotypes with different sequence of defective
alleles. Mixing two or more populations evolving independently decreases the
probability of meeting in one zygote two complementing haplotypes, this
difference results in outbreeding depression (seen in Figs.3b and 4).

\section{Conclusion}
We varied the parameters of the sexual Penna ageing model, in particular by 
separating the population into reproductively isolated groups and/or having
longer bit-strings and a high minimum age of reproduction. We could observe
and interpret inbreeding depression, outbreeding depression, and outbreeding 
advantage, through the counterplay of purifying selection and of haplotype
complementarity. Purifying selection tries to have as few mutations in 
the bit-strings, like haplotype 00000000 for $L=8$, while haplotypes 01100101 
and 10011010 are complementary. In both cases, deleterious effects from
mutations are minimised.

\end{document}